\documentstyle[12pt,epsf]{article}
\begin{document}
\bibliographystyle{unsrt}
\noindent
\begin{center}
{\bf Molecular Dynamics Computer Simulation of the Dynamics of
Supercooled Silica}

J. Horbach, W. Kob\footnote{
Electronic mail: kob@moses.physik.uni-mainz.de\\
http://www.cond-mat.physik.uni-mainz.de/\~{ }kob/home\_kob.html} 
and K. Binder\\
Institute of Physics, Johannes Gutenberg-University, Staudinger Weg 7,\\
D-55099 Mainz, Germany
\end{center}

\vspace*{7mm}
\par
\noindent
\begin{center}
\begin{minipage}[h]{122mm}
We present the results of a large scale computer simulation of
supercooled silica. We find that at high temperatures the diffusion
constants show a non-Arrhenius temperature dependence whereas at low
temperature this dependence is also compatible with an
Arrhenius law. We demonstrate that at low temperatures the intermediate
scattering function shows a two-step relaxation behavior and that it
obeys the time temperature superposition principle. We also discuss the
wave-vector dependence of the nonergodicity parameter and the time and
temperature dependence of the non-Gaussian parameter.
\end{minipage}
\end{center}

\vspace*{5mm}
\par
\noindent

\section{Introduction}
In the last few years ample evidence was given that computer
simulations are a very useful method to study the dynamics of
supercooled liquids (Barrat and Klein 1991, Yip 1995, Kob 1995, Paul
and Baschnagel 1995). Apart from some rare exceptions, such as, e.g.,
the recent simulation of supercooled water by Sciortino {\it et al.}
(Gallo {\it et al.} 1996, Sciortino {\it et al.} 1996), most of these
simulations were done for liquids for which the {\it local} structure is
similar to the one of a closed packed hard sphere system. Thus the
dynamics of systems in which the particles form an open network
structure, such as SiO$_2$ or B$_2$O$_3$, has hardly been investigated
with computer simulations at all, although there is a large body of
experiments in which such materials have been studied (see, e.g., the
papers in Ngai 1994). In this paper we present some of the results of a
large scale molecular dynamics computer simulation in which we
investigated the dynamics of SiO$_2$ in its supercooled state.

\section{Model}

The model we use to describe silica is given by the potential
proposed a few years ago by van Beest {\it et al.} (van Beest, Kramer
and van Santen 1990), and is of the form
\begin{equation}
\phi(r_{ij})=\frac{q_i q_j e^2}{r_{ij}}+A_{ij}e^{-B_{ij}r_{ij}}-
\frac{C_{ij}}{r_{ij}^6}\quad .
\label{eq1}
\end{equation}
The values of the various parameters can be found in the original
publication. This potential has been shown to give a correct
description of the different crystalline phases of silica (Tse and
Klug 1991; Tse, Klug and Allan 1995) and
in a recent publication Vollmayr {\it et al.} (Vollmayr, Kob and Binder
1996) have shown that it is also able to reproduce many static
properties of amorphous silica and that this system forms a network
of tetrahedra, each of which has a silicon atom in its center and which are
interconnected by bridging oxygen atoms. Thus we conclude that this
model gives a quite realistic description of this network glassformer.
Note that in the simulation of Vollmayr {\it et al.} the non-Coulombic
interactions were truncated and shifted at a distance of 5.5\AA $ $  and
in the present work we have done so likewise.

The simulations were done at constant volume and the density of the
system was fixed to 2.3 g/cm$^3$. In order to avoid finite size
effect in the dynamics (Horbach, Kob, Binder and Angell 1996), the
size of the system was relatively large and consisted of 8016 ions.
The equations of motion were integrated with the velocity form of the
Verlet algorithm and the Coulombic contributions to the potential
were calculated with the Ewald summation. The time step of the
integration was 1.6~fs and the longest runs were $4\cdot 10^6$ time
steps, thus 6.4~ns. At each temperature the system was equilibrated
with a stochastic heat bath for a time which was at least as long as
the subsequent production run. The temperatures investigated were
6100~K, 5200~K, 4700~K, 4300~K, 4000~K, 3900~K, 3760~K, 3580~K, 3400~K,
3250~K, 3100~K, 3000~K and 2900 K, the lowest temperature at which the
system could be equilibrated within the time span of our simulation.
More details on the simulation will be given elsewhere (Horbach and Kob
1997).

\section{Results}

\begin{figure}[h]
\caption{Arrhenius plot of the diffusion constant for the silicon
atoms (filled squares) and oxygen atoms (open circles). Thin solid lines: 
Fit with a power-law. Bold solid lines: Fit with an Arrhenius-law.
\label{fig1}}
\caption{Incoherent intermediate scattering function for the silicon
atoms vs. time (a) and rescaled time (b) for all temperatures
investigated.
\label{fig2}}
\caption{Time dependence of the non-Gaussian parameters $\alpha_2$
(a) and $\alpha_3$ (b) for all temperatures investigated (main
figure). Inset: Same data at short times.
\label{fig3}}
\caption{Wave-vector dependence of the nonergodicity parameter for the
incoherent scattering function of the silicon atoms (open circles) and
of the coherent scattering function of the silicon-silicon
correlation (filled squares).
\label{fig4}}
\end{figure}
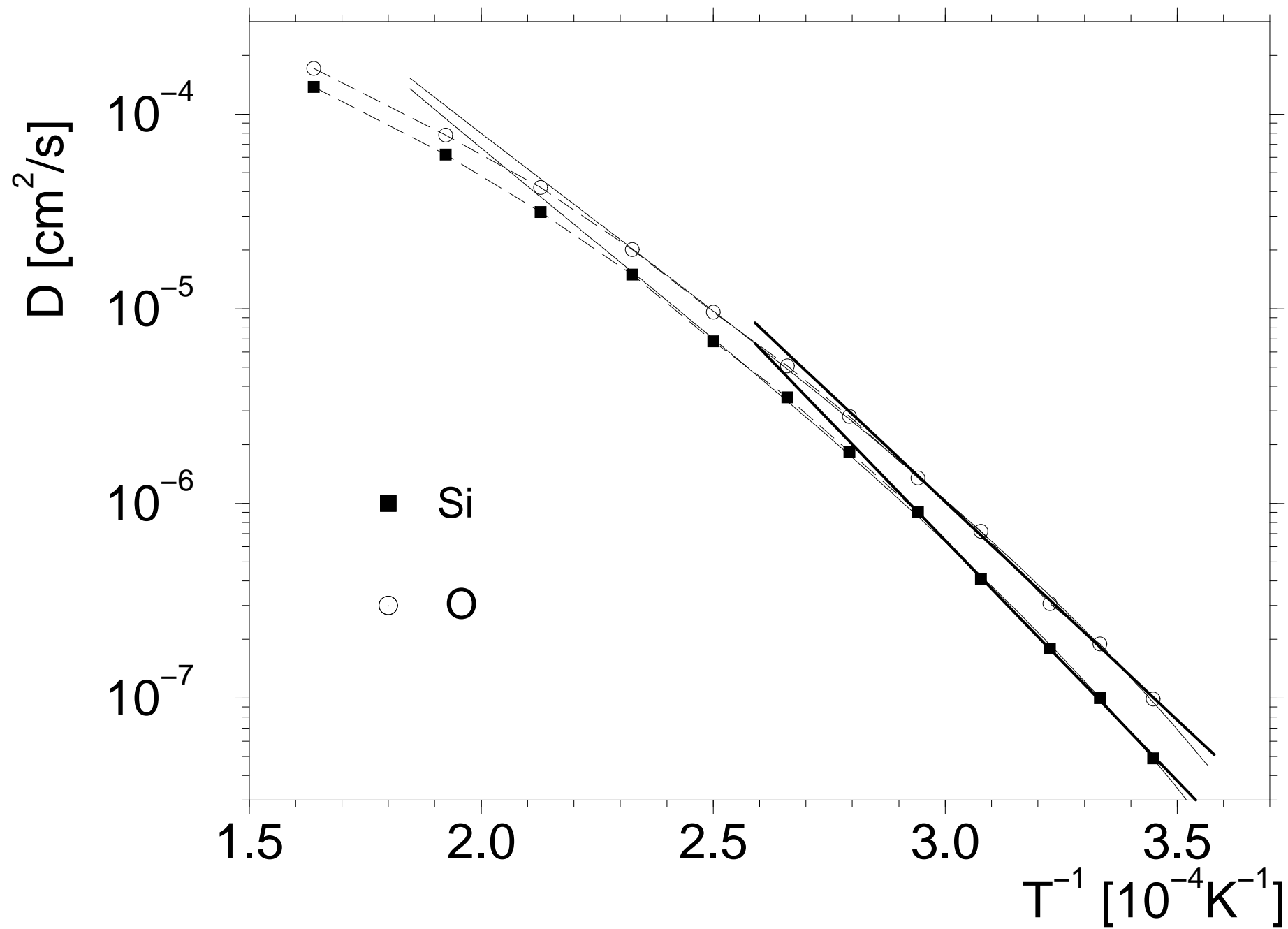
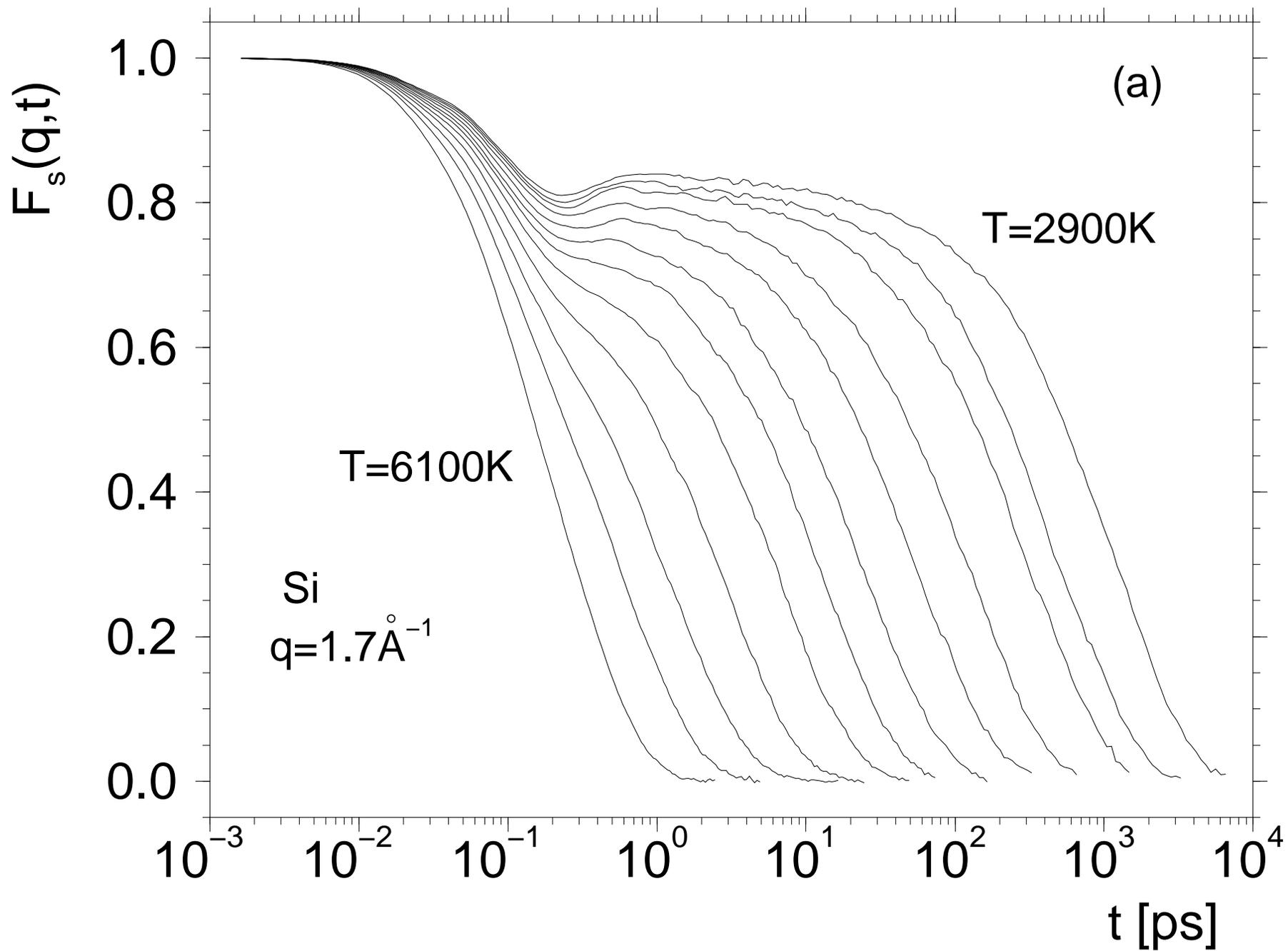
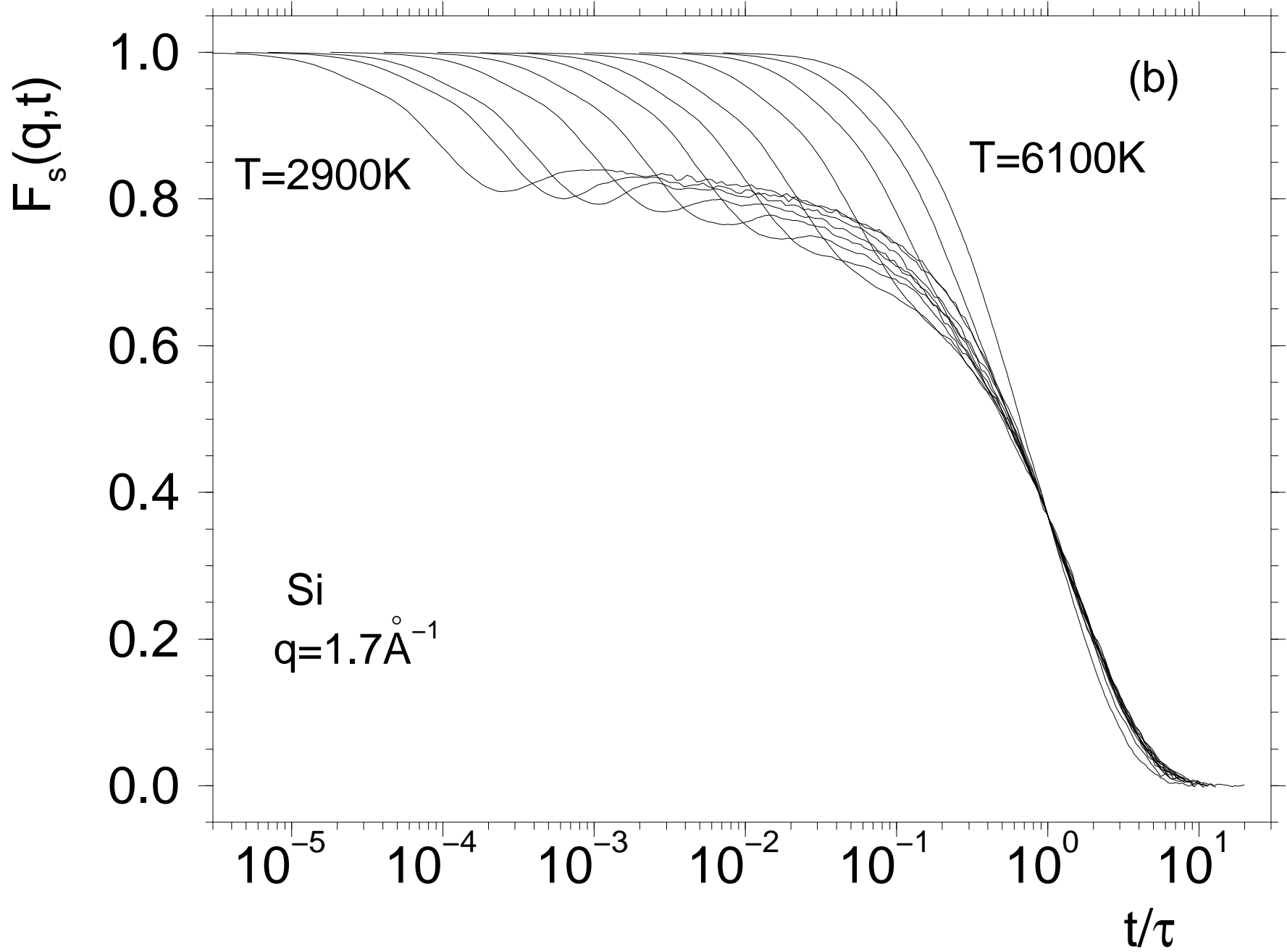
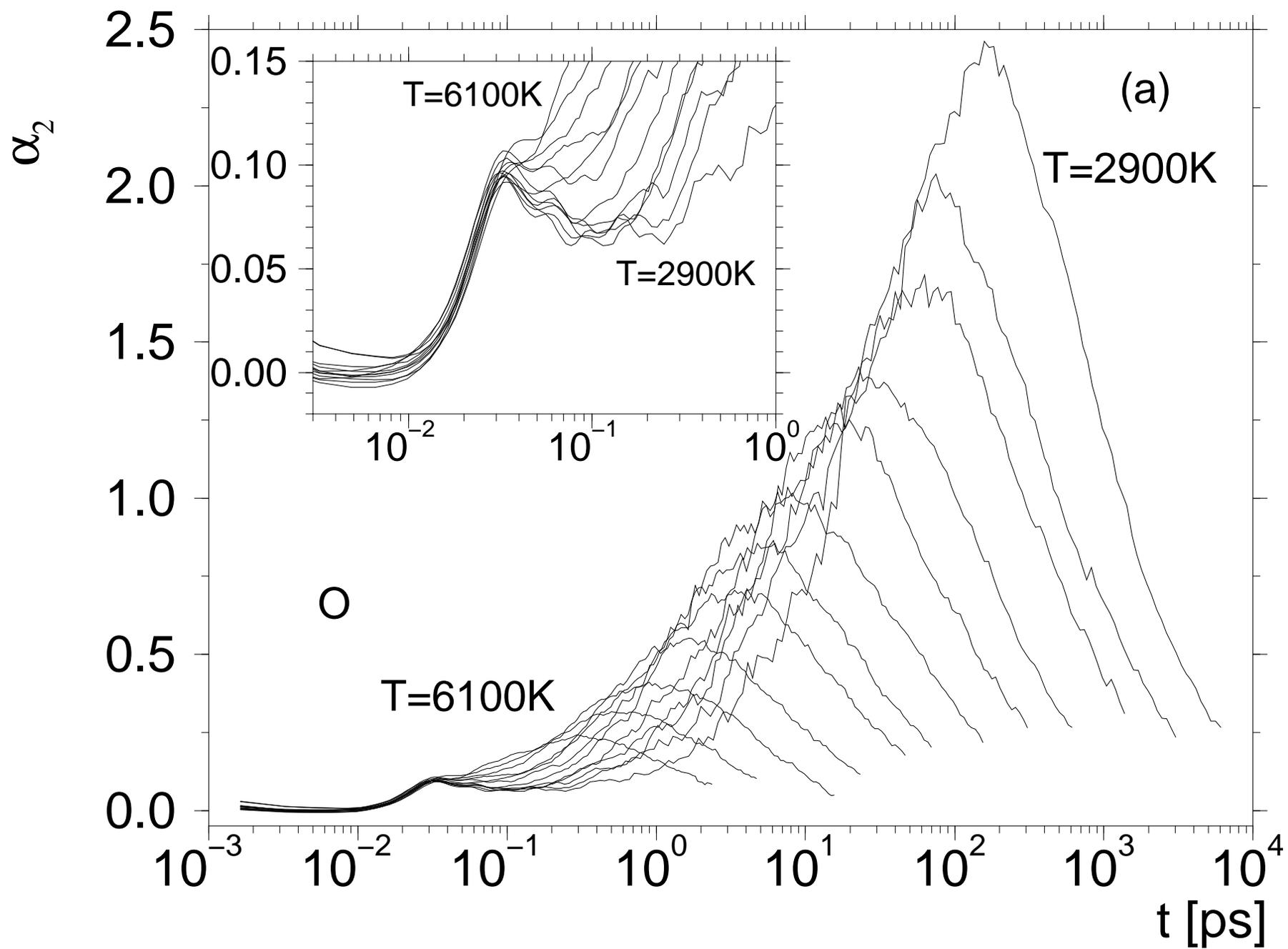
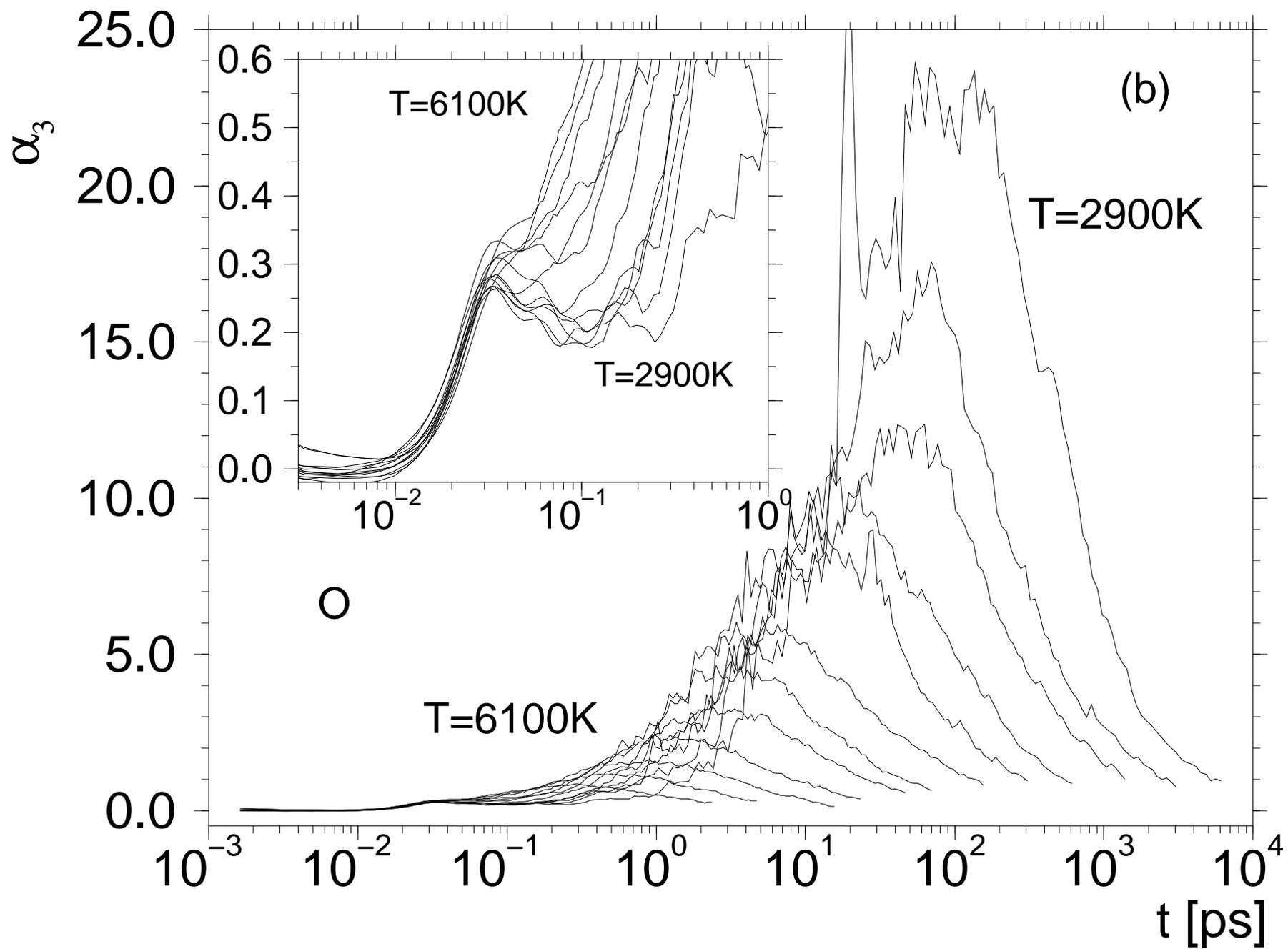
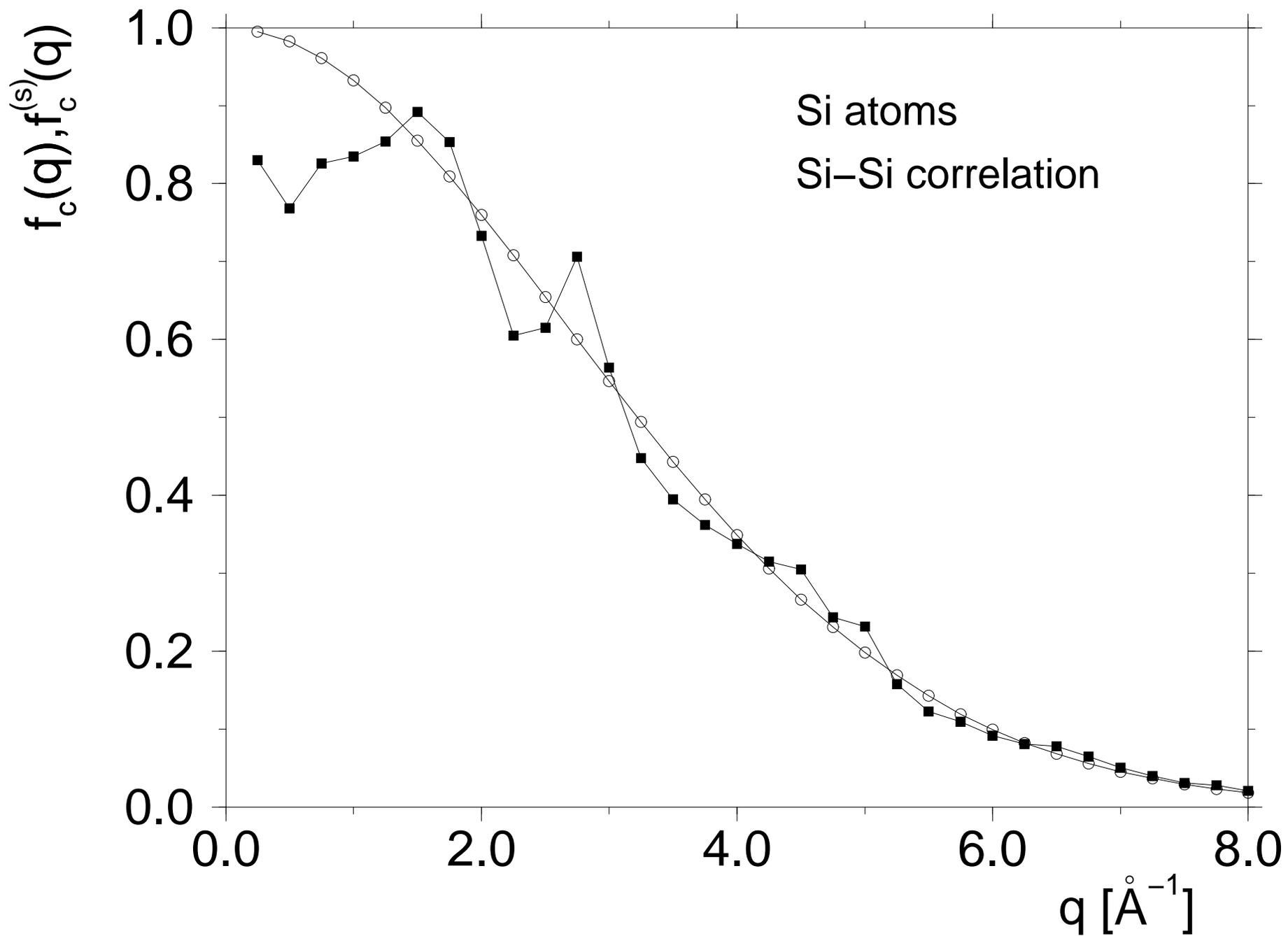

One of the simplest quantities to describe the dynamics of the system
is the tracer diffusion constant $D$ which can be computed from the
mean square displacement of a tagged particle. In Fig.~\ref{fig1} we
show $D$ for the silicon and oxygen atoms in an Arrhenius plot. We
recognize that at high and intermediate temperatures the diffusion
constants clearly show a deviation from the Arrhenius behavior which is
found experimentally, although at lower temperatures than we
investigated (Br\'ebec {\it et al.} 1980, Mikkelsen 1984).  However, at
the lowest temperature considered here, our data is also compatible
with an Arrhenius temperature dependence (bold solid lines). The
activation energies obatined here, 4.45~eV for O and 4.9~eV for Si,
are in surprisingly good agreement with the experimental values
(4.7~eV for O (Mikkelsen 1984) and 6~eV for Si (Br\'ebec
{\it et al.} 1980)).

Motivated by the proposition of the so-called mode-coupling theory (MCT)
(G\"otze 1991, G\"otze and Sj\"ogren 1992), we also attempted to fit
our data for intermediate and low temperatures with a power-law, i.e.
$D \propto (T-T_c)^{\gamma}$. The result of this fit is included in the
figure as well (thin solid lines) and we recognize that this functional
form is able to describe the data very well. Note that the critical
temperature $T_c=2260$ K is the same for both types of atoms. (As one
might have expected, for strong glass formers, $T_c$
is well above the experimental glass transition
temperature $T_g \approx$ 1450 K). Hence we
can infer that at our lowest temperatures the diffusion constants make
a crossover from a power-law behavior to an Arrhenius behavior.  Thus
from our simulation we can make the conjecture that at high enough
temperatures {\it real} silica will make a transition from being a strong
glass former (Angell 1985), which shows an Arrhenius temperature
dependence of the transport coefficients, to a fragile glass former,
which shows a power-law like temperature dependence. This prediction is
also in agreement with recent propositions of R\"ossler and Sokolov
(R\"ossler and Sokolov 1996).

In Fig.~\ref{fig2} we show the time dependence of the incoherent
intermediate scattering function $F_s(q,t)$ for silicon for all temperatures
investigated.  The wave-vector is 1.7 \AA$^{-1}$, the location of the
first sharp diffraction peak in the structure factor. We see,
Fig.~\ref{fig2}a, that even at relatively high temperatures $F_s(q,t)$
shows a shoulder at around 0.6 ps, which develops into a plateau upon
further cooling.  Thus we find that also this network glassformer shows
the typical two-step relaxation behavior that is known for fragile
glasses and whose existence is one of the predictions of MCT. For the
lowest temperatures we also observe that the correlation function shows
a dip around 0.2 ps. This feature is likely related to the so-called
boson peak, which has also been observed in n-scattering experiments
of vitreous silica (Buchenau {\it et al.} 1986).

We define the $\alpha$-relaxation time $\tau(T)$ as the time it takes
the correlation function to decay to $e^{-1}$ of its initial value.
MCT predicts that if a time correlation function is plotted versus the
rescaled time $t/\tau(T)$, one obtains a master curve [time-temperature
superposition principle (TTSP)]. That this is indeed the case is
demonstrated in Fig.~\ref{fig2}b, where we show the same curves as in
Fig.~\ref{fig2}a, but versus $t/\tau(T)$. For rescaled times $t/\tau
\geq 1$, i.e. in the $\alpha$-relaxation regime, the curves for the
different temperatures fall nicely on top of each other, thus showing
that in this time regime the TTSP holds. This is not {\it quite} the
case for $t/\tau \leq 1$, i.e. in the $\beta$-relaxation region.
However, a closer inspection of the individual curves shows that the
reason for the lack of a perfect scaling is likely the fact that the
curves show the aforementioned dip at early times. At the very lowest
temperatures, the location of this dip has been moved to such small
rescaled times, that the scaling of the curves works very well again.
Thus we find that at low enough temperatures the TTSP works also in the
$\beta$-relaxation regime.  Similar conclusions hold also for other
wave-vectors and the oxygen atoms (Horbach and Kob 1997).

The intermediate scattering function can be measured {\it relatively}
easily in scattering experiments. This is not the case for the
so-called non-Gaussian parameters $\alpha_i(t)$, which are a measure
for how strongly $F_s(q,t)$ deviates from a Gaussian behavior (Rahman
1964), whereas these functions can be calculated easily in a computer
simulation.  (In recent n-scattering experiments by Buchenau {\it et
al.} and Zorn the time dependence of $\alpha_2$ was determined
(Buchenau {\it et al} 1996, Zorn 1996)). In Fig.~\ref{fig3} we show the
time dependence of $\alpha_2=3\langle r^4(t)\rangle/5\langle r^2(t)
\rangle^2 -1$ and $\alpha_3 = 3\langle r^6(t) \rangle/35 \langle r^2(t)
\rangle^3 -1$ for the oxygen atoms, were $r(t)$ is the probability that
a tagged particle travels a distance $r$ in time $t$.

From this figure we see that for short times $\alpha_i$ is zero, as is
should be, and then starts to increase when the system enters the
$\beta$-relaxation region. After having reached a maximum value at a
time which corresponds to approximately the end of the
$\beta$-relaxation regime, the curves start to drop back to zero. The
height of this maximum increases with decreasing temperature, thus
showing that the Gaussian approximation breaks down more and more the
lower the temperature is. It is also interesting that for very short
times, $\approx 0.04$ ps, $\alpha_{1/2}$ show a small peak. A similar
peak was also found in a computer simulation of a Lennard-Jones
system (Kob and Andersen 1995a), although with much smaller amplitude.
Since this latter system is not a network forming glass and quite
fragile, we conclude that this feature is probably not closely
connected to the details of the structure of the system but is more
likely a feature that is related to the dynamics of supercooled
liquids.

The last quantity we investigate is the wave-vector dependence of the
nonergodicity parameter, i.e. the height of the plateau in a time
correlation function at low temperatures (G\"otze 1991, G\"otze and
Sj\"ogren 1992). For the case of the coherent and incoherent
intermediate scattering function this is thus nothing else than the
Debye-Waller and Lamb-M\"ossbauer factor, respectively. In
Fig.~\ref{fig4} we show the nonergodicity parameters $f_c^{(s)}(q)$ and
$f_c(q)$ for the incoherent scattering function of the silicon atoms
and the coherent scattering function for the silicon-silicon
correlation, respectively.  We see that $f_c^{(s)}$ shows a Gaussian
like dependence on $q$, as it has already been observed for the case of
simple liquids (Kob and Andersen 1995b), and is also predicted by
MCT.  The nonergodicity parameter for the coherent scattering function
shows several maxima and minima, the location of which are the same as
the ones observed in the structure factor. Also this observation is in
qualitative agreement with the one made for simple liquids and the
prediction of MCT. It is also interesting to note that at a wave-vector
of approximately 8\AA$^{-1}$ the nonergodicity parameters have decayed
to essentially zero, whereas the three partial structure factors show a
significant wave-vector dependence even for larger values of $q$.  This
is in contrast to what has been found in the case of simple liquids
(Kob and Andersen 1995b), where the value of the wave-vector at which
the nonergodicity parameters become zero and the ones at which the
partial structure factors become constant are almost identical. It
would be interesting to understand whether this observation is just a
particularity of silica or whether it is a general property of network
forming glasses.

{\bf Acknowledgements:} This work was supported by BMBF Project 03~N~8008~C
and by SFB 262/D1 of the Deutsche Forschungsgemeinschaft.

\section*{References}
\begin{trivlist}
\item[]
Angell C. A., 1985, 
K. L. Ngai and G. B. Wright (eds.) {\it Relaxation
in Complex Systems}, (US Dept. Commerce, Springfield, 1985).
\item[]
Barrat J.-L., Klein M.L., 1991,
Annu. Rev. Phys. Chem. {\bf 42}, 23.
\item[]
Br\'ebec G., Seguin R., Sella C., Bevenot J., Martin J. C., 1980,
Acta Metall. {\bf 28}, 327.
\item[]
Buchenau U., Prager M., N\"ucker N., Dianoux A. J., Ahmad N.,
Phillips W. A., 1986,
Phys. Rev. B. {\bf 34}, 5665.
\item[]
Buchenau U., Pecharroman C., Zorn R., Frick B., 1996,
Phys. Rev. Lett. {\bf 77}, 659.
\item[]
Gallo P., Sciortino F., Tartaglia P., Chen S.-H., 1996,
Phys. Rev. Lett., {\bf 76}, 2730.
\item[]
G\"otze W., 1991, 
in {\it Liquids, Freezing and the Glass Transition}
Eds.: J. P.  Hansen, D. Levesque and J. Zinn-Justin, Les Houches.
Session LI, 1989, (North-Holland, Amsterdam).
\item[]
G\"otze W., Sj\"ogren L., 1992,
Rep. Prog. Phys. {\bf 55}, 241.
\item[]
Horbach J., Kob W., Binder K., Angell C. A., 1996,
Phys. Rev. E. {\bf 54}, R5889.
\item[]
Horbach J., Kob W., 1997,
(unpublished).
\item[]
Kob W., 1995,
p. 1 in Vol. III of {\it Annual Reviews of Computational Physics},
Ed.: D. Stauffer (World Scientific, Singapore, 1995).
\item[]
Kob W., Andersen H. C., 1995a,
Phys. Rev. E {\bf 51}, 4626.
\item[]
Kob W., Andersen H. C., 1995b,
Phys. Rev. E {\bf 52}, 4134.
\item[]
Mikkelsen J. C., 1984,
J. Appl. Phys. Lett. {\bf 45}, 1187.
\item[]
Ngai K.L., 1994,
Ed.: {\it Proc. 2nd International Discussion Meeting on
Relaxations in Complex Systems}, J. Non-Cryst. Solids {\bf 172-174}.
\item[]
Paul W., Baschnagel J., 1995,
in {\em Monte Carlo and Molecular Dynamics Simulations in Polymer Science}, 
Ed.: K. Binder, (Oxford University Press, New York).
\item[]
Rahman A., 1964,
Phys. Rev. {\bf 136}, A 405.
\item[]
R\"ossler E., Sokolov A. P., 1996,
Chemical Geology, {\bf 128}, 143.
\item[]
Sciortino F., Gallo P., Tartaglia P., Chen S.-H., 1996,
Phys. Rev. E, {\bf 54}, 6331.
\item[]
Tse J. S., Klug D. D., 1991,
Phys. Rev. Lett. {\bf 67}, 3559.
\item[]
Tse J. S., Klug D. D., Allan D. C., 1995,
Phys. Rev. B {\bf 51}, 16392.
\item[]
van Beest B. W., Kramer G. J., van Santen R. A., 1990,
Phys. Rev. Lett.  {\bf 64} 1955.
\item[]
Vollmayr K., Kob W., Binder K., 1996,
Phys. Rev. B {\bf 54}, 15808.
\item[]
Yip S., 1995, 
Theme Issue on Relaxation Kinetics in Supercooled Liquids-Mode Coupling
Theory and its Experimental Tests; Ed. S. Yip. Volume {\bf 24}, No.
6-8 of {\it Transport Theory and Statistical Physics}.
\item[]
Zorn R, 1997,
Phys. Rev. B., {\bf 55}, xxxx.

\end{trivlist}

\section*{Figures}
\end{document}